\documentclass[aps,pra,twocolumn,nofootinbib,longbibliography,10pt]{revtex4-2}
\usepackage{hyperref}
\usepackage{graphicx} 
\usepackage{amsmath}
\usepackage{amsfonts}
\usepackage{amssymb}
\usepackage{graphicx}
\usepackage{dcolumn}
\usepackage{bm}
\usepackage{hyperref,etoolbox}
\usepackage{mathrsfs}
\usepackage{subfig}
\usepackage{natbib}
\usepackage{caption}
\usepackage[none]{hyphenat}

\begin{document}

\title{\bf Entropic force for quantum particles}
\author{\textbf{Jayarshi Bhattacharya}}
\email{dibyajayarshi@gmail.com}
\author{\textbf{Gautam Gangopadhyay}}
\email{gautam@bose.res.in}
\author{\textbf{Sunandan Gangopadhyay}}
\email{sunandan.gangopadhyay@gmail.com}\thanks{(Corresponding Author)}
\affiliation{\textit{S.N. Bose National Centre for Basic Sciences, JD Block, Sector-III, Salt Lake, Kolkata 700106, India}}
\begin{abstract}
\noindent Entropic force has been drawing the attention of theoretical physicists following E. Verlinde's  work in 2011 to derive Newton's second law and Einstein's field equations of general relativity. In this paper, we extend the idea of entropic force to the distribution of quantum particles. Starting from the definition of Shannon entropy for continuous variables, here we have derived quantum osmotic pressure as well as the consequent entropic forces for bosonic and fermionic particles. The entropic force is computed explicitly for a pair of bosons and fermions. The low temperature limit of this result show that the entropic force for bosons is similar to Hooke's law of elasticity revealing the importance of this idea in the formation of a Bose-Einstein condensate. For fermions, the low temperature limit boils down to the well known Neumann's radial force and also reveals the Pauli's exclusion principle. The classical limit of the entropic force between quantum particles is then discussed. As a further example, the entropic force for quantum particles in noncommutative space is also computed. The result reveals a violation of the Pauli exclusion principle for fermions in noncommutative space.\\
\noindent DOI \href{https://iopscience.iop.org/article/10.1088/1402-4896/ace389}{10.1088/1402-4896/ace389}
\end{abstract}
\maketitle
\section{Introduction}
An entropic force is created due to the nature of thermodynamic systems to maximise its entropy. It is an effective macroscopic force resulting from the microscopic elements in a system with many degrees of freedom which has statistical tendency to increase its entropy. In general, the system, or the set of macroscopic variables such as pressure, volume or the entropic force as well, evolves towards the statistically more probable state from the less probable state \cite{muller2007history}. The way the system evolves appears to be indeed a force. We can consider diffusion as an example, where the diffusion pressure leads the process.\\
A few years back, a different approach to reach Newton's second law of motion was proposed from the definition of entropic force \cite{verlinde2011origin}. This new idea could also reproduce a part of the Einstein's field equations of general relativity \cite{einstein1917special}.\\
In \cite{roos2014entropic}, it was shown that the Hooke's elasticity law can be derived from the idea of entropic force. The expression for osmotic forces for the Brownian particles was also established from the definition of entropic force. Some recent works on entropic forces of quantum particles have appeared in \cite{mehdipour2015entropic},\cite{shah2014quantum}. These findings indicate that the idea of entropic force is something very fundamental.\\
In this paper we try to generalise the approach in \cite{roos2014entropic} further. We establish the Einstein relation for the osmotic force \cite{einstein1905molekularkinetischen} starting from the Shannon entropy for continuous variables \cite{caticha2012entropic}. Maximising this entropy leads to the Maxwell-Boltzmann distribution. Taking a gradient of this leads to the Einstein relation. We then extended this idea to the quantum domain replacing the classical particle distribution with Bose-Einstein and Fermi-Dirac particle distributions.\\
The paper is organised as follows. In section \ref{sec1} we discuss about continuous entropy and how it can lead to the entropic force. We establish an identity between entropic force and classical force. In section \ref{sec2}, we generalise this idea to the quantum regime and obtained deviations from the classical result. In section \ref{sec3}, we compute the quantum entropic force in noncommutative space. Finally, we conclude in section \ref{sec4}.


\section{Equivalence of entropic force and classical force}{\label{sec1}}
Entropy is a fundamental concept of physics. From the second law of thermodynamics, entropy is considered to be a measure of randomness in a system. The measure of randomness is calculated by the number of microstates and connecting it to Boltzmann's law. This measure of microstates can be resembled through the eye of information theory considering entropy to be expected value of missing information.\\
When the probability distribution is continuous, we can define a probability density function and similarly the idea of entropy becomes continuous. In \cite{marsh2013introduction},\cite{mohtashami2019difference}, it was shown that Shannon entropy, can be written in continuous form with some restrictions applied. By the mean value theorem, a continuous distribution $f$ can be discretized into bins of size $\Delta$. This implies that form $x_i$ to $x_i+\Delta$, $f$ can be approximated by as
\begin{equation}\label{0a}
    \lim_{\Delta \to 0} \int_{x_i}^{x_i+\Delta}f(x) dx \rightarrow f(x_i)\Delta ~.
\end{equation}
However, in this description, there is a difference between continuous entropy and Shannon entropy given by \cite{shannon1948mathematical}
\begin{equation}\label{0b}
    S=-\sum_{i} p_i\;\ln\; p_i~.
\end{equation}
Now, we can consider a probability density function $p(x)$, such that each bean gives the probability
\begin{equation}
    p_i=p(x_i)\Delta
\end{equation}
Then eq.\eqref{0b} can be approximated in a Riemann sum
\begin{align}
    S_\Delta &= -\lim_{\Delta \to 0}\sum_{i=-\infty}^{+\infty} p(x_i)\Delta \ln \left( p(x_i)\Delta \right)\nonumber\\
    &=-\int_{-\infty}^{+\infty}p(x)\ln\; p(x)  dx - \lim_{\Delta \to 0}\sum_{i=-\infty}^{+\infty} p(x_i)\Delta \ln\Delta~.
\end{align}
The second term on the right hand side of the above equation blows up as $\ln(\Delta)\rightarrow -\infty$ when $\Delta \rightarrow 0$. Hence, continuous entropy is defined as \cite{marsh2013introduction}
\begin{align}\label{rev1}
    S &= \lim_{\Delta \to 0} \left(S_\Delta + \sum_{i=-\infty}^{+\infty} p(x_i)\Delta \ln\Delta\right)\nonumber\\
    &=-\int_{-\infty}^{+\infty}p(x)\ln p(x)  dx~.
\end{align}
This is called differential entropy and from the above analysis it is clear that this form of the entropy does not follow from the limit of the discrete case \cite{mohtashami2019difference}. But even this definition of continuous entropy is not scale invariant. Indeed if a change of variable is made as $Y=\alpha X$, $\alpha$ being a constant, the entropy $S(X)$ changes,  
where $S(X)$ is given by
\begin{align}\label{rev1aa}
S(X)=-\int_{-\infty}^{+\infty}p_X(x)\ln p_X(x)dx ~. 
\end{align}
To see this, we first note that 
\begin{align}\label{rev1ab}
\left|p_Y(y)dy\right|=\left|p_X(x)dx\right|
\end{align}
following from the fact that the probability contained in an infinitesimal area should be invariant under a change of variables. Then we have
\begin{align}\label{rev2}
	S(Y)&=-\int_{-\infty}^{+\infty}p_Y(y)\ln p_Y(y)dy \nonumber\\
		&=-\int_{-\infty}^{+\infty}\;p_X(x)\ln\left(p_X(x)\left|\frac{dx}{dy}\right|\right)dx\nonumber\\
		&=-\int_{-\infty}^{+\infty}p_X(x)\ln p_X(x)dx -\int_{-\infty}^{+\infty}\;p_X(x)\ln\left|\frac{dx}{dy}\right|dx\nonumber\\
		&=S(X)-\int_{-\infty}^{+\infty}\;p_X(x)\ln\left|\frac{dx}{dy}\right|dx\nonumber\\
		&=S(X)+\ln|\alpha|\int_{-\infty}^{+\infty}\;p_X(x)dx \nonumber\\
		&=S(X)+\ln|\alpha|
\end{align}
where in the second line we have used eq.\eqref{rev1ab}, in the fourth line we have used eq.\eqref{rev1aa}, in the penultimate line we have used $y=\alpha x$, and in the last line we have used
\begin{align}\label{rev2z}
\int_{-\infty}^{+\infty}\;p_X(x)dx=1~.
\end{align}
To resolve this problem, the idea of relative entropy is introduced \cite{marsh2013introduction}. The  relative entropy is defined as
\begin{equation}\label{rev3}
	S(p||q)=-\int_{-\infty}^{+\infty}p(x)\ln\left(\frac{p(x)}{q(x)}\right)dx~.
\end{equation}
Following the steps leading to eq.\eqref{rev2}, it can be shown that the above form of the entropy is invariant under change of variables.

\noindent In three dimensions, the definition of entropy reads
\begin{equation}
    S\;=\;-\;k_B\int p(\vec{r}) \ln\left(\frac{p(\vec{r})}{q(\vec{r})}\right) d^3\vec{r}\label{1}~.
\end{equation}
In the subsequent discussion, we now choose $q(\vec{r})$ to be a uniform distribution, that is,  $q(\vec{r})=1/V$, $V=\int d^3 \vec{r}$.

\noindent Now we move on to maximise the above expression to get the probability distribution where $p(\vec{r})$ is the probability distribution of the system satisfying
\begin{equation}
    \int p(\vec{r})\;d^3\vec{r}\;=\;1\label{2}~.
\end{equation}
The average energy of the system can be written as
\begin{equation}
    \left<U(\vec{r})\right>\;=\;\int p(\vec{r})U(\vec{r})\;d^3\vec{r} \label{3} ~.
\end{equation}
To maximise the entropy, we need to incorporate the constraints \eqref{2}, \eqref{3} through Lagrangian multipliers $\lambda$ and $\beta$
\begin{align}\label{4}
    \frac{S}{k_B} = &-\int p(\vec{r}) \ln\left(p(\vec{r}V)\right) d^3\vec{r} -\lambda\left(\int p(\vec{r})\;d^3\vec{r}-1\right) \nonumber\\
    &-\beta\left(\int p(\vec{r})U(\vec{r})\;d^3\vec{\vec{r}}-\left<U(\vec{r})\right>\right)~.
\end{align}
Maximising the entropy $S$ in eq.\eqref{4}, we get
\begin{align}
    0&=\delta\left(\frac{S}{k_B}\right)\nonumber\\
     &= -\int d^3\vec{r}\left[\ln\left(p(\vec{r})V\right)+1+\lambda+\beta U(\vec{r})\right]\delta p(\vec{r})\label{4a}~.
\end{align}
As $\delta p(\vec{r})$ is arbitrary for all $\vec{r}$, from eq.\eqref{4a} we have
\begin{equation}\label{4b}
	\ln\left(p(\vec{r})V\right)+1+\lambda+\beta U(\vec{r}) =0~.
\end{equation}
This then leads to
\begin{equation}\label{5}
    p(\vec{r})=\frac{1}{Q} e^{-\beta U(\vec{r})}
\end{equation}
where $Q$ is a constant given by
\begin{equation}\label{6}
    Q= V e^{1+\lambda}=\int e^{-\beta U(\vec{r})} d^3\vec{r}
\end{equation}
where the second equality follows from eq.\eqref{2}.

\noindent We now take the gradient of the probability to get an important result. From eq.\eqref{5} we can write
\begin{equation}\label{7}
    -\vec{\nabla} U(\vec{r}) = \frac{1}{\beta}\frac{\vec{\nabla} p(\vec{r})}{p(\vec{r})}
\end{equation}
where $\beta=\frac{1}{k_BT}$ from thermodynamics \cite{10.5555/1996289}, $T$ being the temperature of the system.\\
We now look at eq.\eqref{7} carefully. From classical mechanics \cite{goldstein:mechanics}, we have $- \vec{\nabla} U(\vec{r})=\Vec{F}_c (\vec{r})$, where $\Vec{F}_c(\vec{r})$ is a classical force \footnote{From now on we shall assume the potential $U(r)$ to be spherically symmetric which results in a spherically symmetric probability density function.}.\\
So we can recast eq.\eqref{7} as
\begin{equation}\label{8}
    \Vec{F}_c (r) = k_B T \frac{\vec{\nabla} p(r)}{p(r)}~.
\end{equation}
Without a negative sign, the above equation resembles Einstein's famous definition of osmotic force \cite{einstein1905molekularkinetischen}
\begin{equation}\label{9}
    \Vec{F}_{osm} (r) = -k_B T \frac{\vec{\nabla} p(r)}{p(r)}~.
\end{equation}
This is a very general result and was obtained in \cite{roos2014entropic} in the context of Brownian particles. However, in general the result holds for any Maxwell-Boltzmann distribution.\\
We now briefly clarify the formalism to define the entropic force \cite{verlinde2011origin}. Considering a system in the microcanonical ensemble with an external applied force, we let the total energy $E$ be varied to $E+F_e\cdot r$. Then maximising the entropy for the system, one can write \cite{verlinde2011origin, roos2014entropic}
\begin{align}
    \hspace{2cm}\frac{\partial}{\partial r}S(E+F_e\cdot r) &=0 ~.
\end{align}
Carrying out a Taylor series expansion gives
\begin{align}
    \frac{\partial}{\partial r}\left\{S(E)+r\cdot F_e\frac{\partial S}{\partial E}+O(r^2)\right\}=0~.
\end{align}
This implies
\begin{align}
    \hspace{1.5cm}F_e (r) = - T\frac{\partial S}{\partial r}  \label{scalar}
\end{align}
where $T^{-1}=\frac{\partial S}{\partial E}$ is defined to be the inverse of the temperature. Hence one can write eq.\eqref{scalar} in vector notation \footnote{The negative sign in eq.\eqref{scalar} differs from the original result in \cite{verlinde2011origin}. But we shall see in the subsequent discussion that this sign will be crucial in interpreting the results that we obtain for the entropic force between quantum particles.}
\begin{equation}
    \vec{F}_e (r) = - T\vec{\nabla} S  \label{10}
\end{equation}
From statistical mechanics \cite{Huang_1987}, we know that Boltzman entropy is given by
\begin{equation}
    S=k_B\; \ln \;\Omega \label{11}
\end{equation}
and if the microstates have equal appriori probability, then we can write
\begin{equation}
    \Omega=\frac{1}{p(r)}\label{12}~.
\end{equation}
Thus using eq.(s) (\eqref{11}, \eqref{12}) in eq.\eqref{10}, we can write
\begin{align}
    \vec{F}_e (r) &=- T\;\vec{\nabla} \left(k_B \ln \Omega\right) \nonumber\\
             &=-k_B T\;\vec{\nabla} \left(\ln \frac{1}{p(r)}\right) \nonumber\\
             &=k_B T \frac{\vec{\nabla} p(r)}{p(r)} \nonumber\\
             &=-\vec{F}_{osm} (r)\label{13}~.
\end{align}
This implies from eq.(s) (\eqref{8},\eqref{13}) that in general we can write 
\begin{equation}
    \vec{F}_c=\vec{F}_e \label{ex1}~.
\end{equation}
Hence, the classical force and the entropic force for classical particles are exactly equal.


\section{Quantum entropic force}{\label{sec2}}
It is evident that we have built the idea of entropic force from a classical distribution of the particles. We would now like to extend this idea to quantum distribution of particles. We would like to apply the definition of entropic force for quantum particles such as bosons and fermions. From the concept of density matrix for free particle in quantum statistical mechanics \cite{Huang_1987},\cite{pathria1996statistical}, the density matrix in coordinate representation reads
\begin{align}
    \hat{\rho}=\frac{e^{-\beta\hat{H}}}{tr\left(e^{-\beta\hat{H}}\right)}
    \label{d10}
\end{align}
where $\hat{H}$ represents the free particle Hamiltonian
\begin{align}
    \hat{H}=\frac{\hat{p}_1^2}{2m}+\frac{\hat{p}_2^2}{2m}~.
    \label{d12}
\end{align}
$\hat{p}_1$ and $\hat{p}_2$ denote the momentum operators of the particles. 

\noindent The probability density for the two particles to be at positions $\vec{r}_1$ and $\vec{r}_2$ can be written as \cite{Huang_1987},\cite{pathria1996statistical}
\begin{align}           
p(\Vec{r}_1,\Vec{r}_2)&=\left<\vec{r_1},\vec{r_2}\mid\hat{\rho}\mid\vec{r_1},\vec{r_2}\right>\nonumber\\
    &= \frac{\left<\vec{r_1},\vec{r_2}\mid e^{-\beta\hat{H}}\mid\vec{r_1},\vec{r_2}\right>}{tr\left(e^{-\beta\hat{H}}\right)}\nonumber\\
    &\approx \frac{1}{V^2}\left(1\pm exp\left(-\frac{2\pi r_{12}^2}{\lambda_{th}^2}\right)\right) \label{15}
\end{align}
where $\vec{r}_{12}=\vec{r}_1-\vec{r}_2$, $V$ denotes the volume of the chamber that contains the bosons or fermions and $\lambda_{th}=\hbar\sqrt{\frac{2\pi\beta}{m}}$ is the thermal wavelength. The $\pm$ signs denotes the Bose-Einstein and Fermi-Dirac statistics respectively.\\
We write this two particle probability density function as
\begin{align}
    p(r) &= \frac{1}{V^2}e^{-\beta v_s(r)} \label{16}
\end{align}
where $v_s$ is defined to be statistical interparticle potential given by
\begin{align}
    v_s(r) &= -k_B T \ln\left(1\pm exp\left(-\frac{2\pi r^2}{\lambda_{th}^2}\right)\right)~. \label{17}
\end{align}
We can now define the quantum entropic force for this scenario using the same definition given by eq.\eqref{13}. With the form of $p(r)$ given in eq.\eqref{16}, we can now define the quantum entropic force as
\begin{align}
    \vec{F}_e^ q(r) &=k_B T \frac{\vec{\nabla} p(r)}{p(r)}\nonumber\\
            &= -\vec{\nabla} v_s(r)\label{18}\\
            &= \vec{F}_s(r)\nonumber~.
\end{align}
Hence, we have the relation $\vec{F}_s(r)=\vec{F}_e^ q(r)$, where $\Vec{F}_s(r)$ is the statistical force. Interestingly, the statistical force $\vec{F}_s(r)$ and the quantum entropic force $\vec{F}_e^ q(r)$ for quantum particles like bosons and fermions are also equal.\\
From eq.(\eqref{17},\eqref{18}) we can write the quantum osmotic force to be
\begin{align}
    \vec{F}_e^ q(r) &= -\frac{4\pi}{\lambda_{th}^2\beta}\frac{r}{1\pm exp\left(\frac{2\pi r^2}{\lambda_{th}^2}\right)}\hat{e}_r\nonumber\\
    &= -\frac{2m k_B^2 T^2}{\hbar^2}\frac{r}{1\pm exp\left(\frac{m k_B T r^2}{\hbar^2}\right)}\hat{e}_r\label{19}
\end{align}
where we have used the expression $\lambda_{th}=\hbar\sqrt{\frac{2\pi\beta}{m}}$ and $\beta=\frac{1}{k_B T}$ in the second line of the above equation. Here $\hat{e}_r=\frac{\Vec{r}}{r}$ denotes the unit vector and the $\pm$ sign denotes bosons and fermions.\\
\begin{figure}[h]
    \centering
    \includegraphics[width=0.5\textwidth]{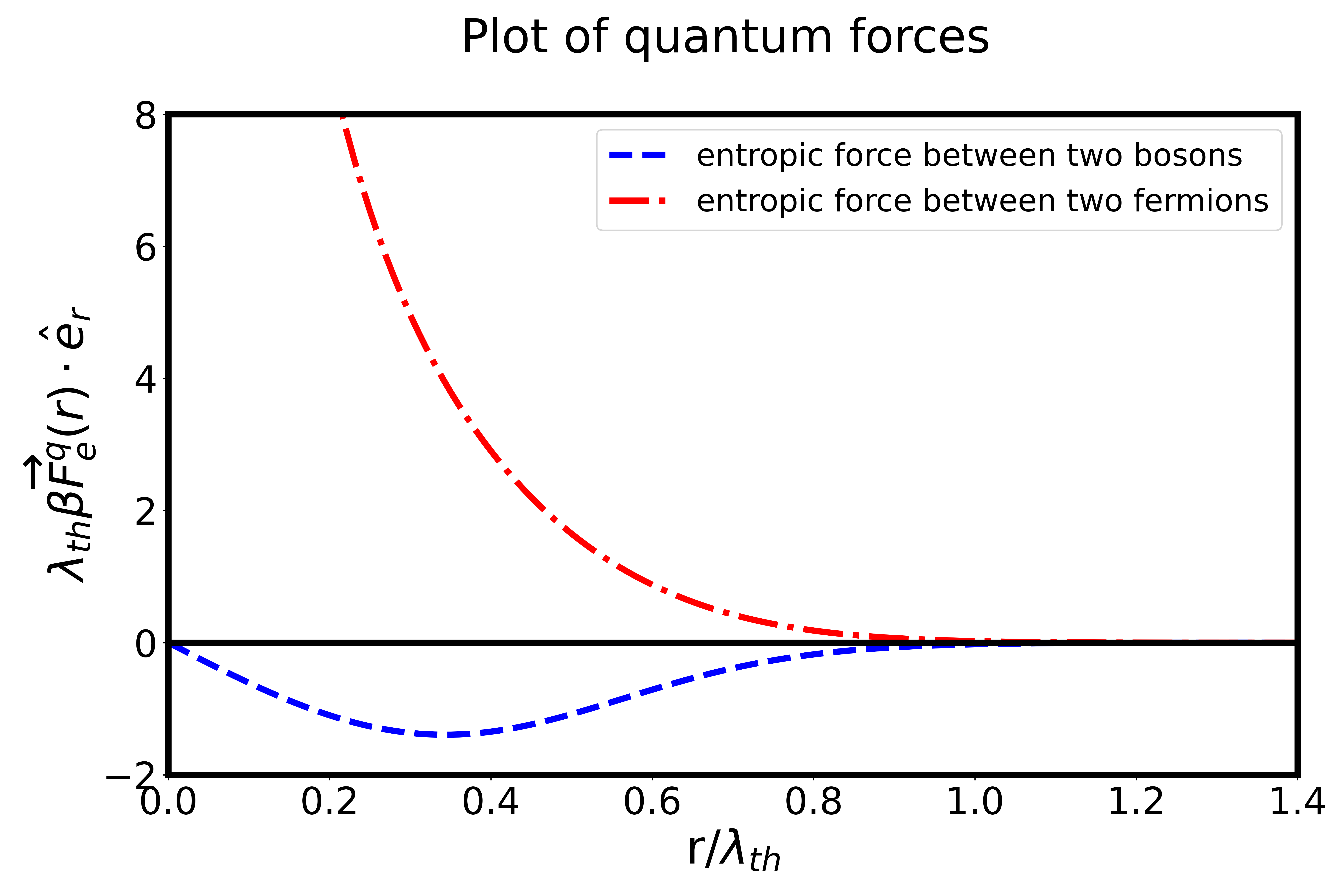}
    \caption{Interparticle entropic force (as $\lambda_{th}\beta \vec{F}_e^q(r)\cdot\hat{e}_r$) is plotted against the separation (as $r/\lambda_{th}$) for bosons and fermions.}
    \label{fig:my_label}
\end{figure}
\\We shall now see that the low temperature limit of the above result gives rise to some more interesting results.
\subsection{Low temperature limit}{\label{sub1}}
At low temperature, the particles will show quantum properties. But if the particles are enough separated, the quantum effect will be diluted. So we focus on a situation when the particles are close enough and the temperature is very low, that is, $\frac{m k_B T r^2}{\hbar^2}<<1$. In this case the quantum entropic force for bosons will be
\begin{equation}
    \vec{F}_e^ q(r) \approx -\frac{2\pi k_BTr}{\lambda_{th}^2}\hat{e}_r\equiv-a\Vec{r} \label{20}
\end{equation}
where $a= \frac{2\pi k_BT}{\lambda_{th}^2}$.\\
This is a new and interesting result which suggests that the entropic force for bosons will increase in magnitude with separation. The negative sign denotes the attractive nature between two bosons and reveals that entropic force plays a key role in driving the bosons to form a Bose-Einstein condensate \cite{einstein1924quantum}.
 The form of the entropic force also looks similar to the Hooke's law of elasticity. The idea of Hooke's law comes from a very simple model of a polymer consisting of $N_{ch}$ rigid monomers of length $l_{ch}$ whose orientations are independent of each other. The chain forms a pattern similar to that of a random walk. Hence, when one end of the chain is fixed to the origin, the probability distribution for the other endpoint $r$ of the chain is given by
\begin{equation}
    p(r) = \left(\frac{3}{2\pi^2 N_{ch} l_{ch}^2}\right)^{\frac{3}{4}}exp\left(-\frac{3r^2}{2 N_{ch} l_{ch}^2}\right)~.
\end{equation}
This leads to a force for a chain embedded in a medium with temperature $T$ of the form
\begin{align}
    \Vec{F}_{ch} &= k_BT\Vec{\nabla}\ln p(r) \nonumber\\
                 &= -k_BT\frac{3r}{N_{ch} l_{ch}^2}\hat{e}_r\nonumber\\
                 &= -a\Vec{r} \label{extra2}~.
\end{align}
For fermions, the low temperature limit of the quantum entropic force reads
\begin{align}
    \vec{F}_e^ q(r) \approx \frac{2 k_B T}{r}\hat{e}_r~.\label{21}
\end{align}
This is another key result for forces between fermionic particles. The positive sign shows the repelling nature between two fermions and also shows that the entropic force blows up at $r=0$ revealing Pauli's exclusion principle. This result also looks exactly identical in magnitude to the Neumann's radial force for Brownian particles \cite{neumann1980entropic}. As explained in \cite{roos2014entropic}, for a Brownian particle which has moved to a distance $r$ from a fixed centre, the available volume increases as $4\pi r^2dr$ and the number of microstates $\Omega$ therefore scales as $r^2$, which in turn leads to the Neumann’s radial force \cite{neumann1980entropic}
\begin{align}
    \Vec{F}_r &= \frac{2k_B T}{r}\hat{e}_r \label{extra1}~.
\end{align}
\subsection{Classical limit}{\label{sub2}}
We now discuss the classical limit. The classical limit takes over when the temperature is high or the particles are separated enough that is $r>>\lambda_{th}$. In both cases, $\frac{m k_B T r^2}{\hbar^2}>>1$ which implies that $e^{-\frac{m k_B T r^2}{\hbar^2}}<<1$. Then the entropic force is given by
\begin{align}
    \vec{F}_e^ q(r) &= -\frac{2 k_B T}{r}\frac{\frac{m k_B T r^2}{\hbar^2}}{1\pm exp\left(\frac{m k_B T r^2}{\hbar^2}\right)}\hat{e}_r\nonumber\\
    &=\mp\frac{2 k_B T}{r}\frac{m k_B T r^2}{\hbar^2}e^{-\frac{m k_B T r^2}{\hbar^2}} \left(1\pm e^{-\frac{m k_B T r^2}{\hbar^2}}\right)\hat{e}_r\nonumber\\
    &\approx \mp \frac{2\pi k_BTr}{\lambda_{th}^2} e^{-\frac{2\pi r^2}{\lambda_{th}^2}} \left(1\mp O\left(e^{-\frac{2\pi r^2}{\lambda_{th}^2}}\right)\right)\hat{e}_r \nonumber\\
     &\approx \frac{2\pi k_BTr}{\lambda_{th}^2} \left(\mp e^{-\frac{2\pi r^2}{\lambda_{th}^2}}+ O\left(e^{-\frac{4\pi r^2}{\lambda_{th}^2}}\right)\right)\hat{e}_r~. \label{22}
\end{align}
Hence for bosons, the force is,
\begin{equation}
    \vec{F}_e^ b(r) =- \frac{2\pi k_BTr}{\lambda_{th}^2} \left(e^{-\frac{2\pi r^2}{\lambda_{th}^2}}- e^{-\frac{4\pi r^2}{\lambda_{th}^2}}+ ...\right)\hat{e}_r \label{23}
\end{equation}
and for fermions, the force is,
\begin{equation}
    \vec{F}_e^ f(r) = \frac{2\pi k_BTr}{\lambda_{th}^2} \left(e^{-\frac{2\pi r^2}{\lambda_{th}^2}}+ e^{-\frac{4\pi r^2}{\lambda_{th}^2}}+ ...\right)\hat{e}_r~. \label{24}
\end{equation}
The opposite signs denote the attractive nature of force for bosons and the repulsive nature for fermions. Both the forces go to zero as the distance between the two particles go to infinity.

\section{Quantum entropic force in noncommutative space}{\label{sec3}}

\noindent In this section, we look at another example where we calculate the entropic force. In the study of  noncommutative geometry \cite{snyder1947quantized}, \cite{seiberg1999string}, the two particle correlation function defined in section \ref{sec2}, takes an interesting form. 
The idea of noncommutative spacetime was introduced in \cite{snyder1947quantized} back in 1947 in order to control ultra-violet divergences. However, it was not taken seriously till recently when noncommutativity of spacetime coordinates emerged as a consequence of studies
in string theory \cite{seiberg1999string}.

\noindent In \cite{chakraborty2006twisted}, it was shown that for the commutation relation between spatial coordinates $\hat{x}$ and $\hat{y}$
\begin{align}\label{addition1}
    [\hat{x},\hat{y}]=i\theta
\end{align}
 where $\theta$ is the noncommutative parameter, the two particle correlation function for bosons and fermions becomes
 \begin{align}\label{addition2}
     p(r) &= \frac{1}{V^2}\left(1\pm \frac{exp\left(-\frac{2\pi r^2}{\lambda_{th}^2(1+\frac{\theta^2}{\lambda_{th}^4})}\right)}{1+\frac{\theta^2}{\lambda_{th}^4}} \right)
 \end{align}
 where the $\pm$ signs denote the Bose-Einstein and Fermi-Dirac statistics respectively. $V$ and $\lambda_{th}$ has the same meaning as earlier. Then the entropic force, defined in eq.\eqref{18} takes the form
\begin{equation}\label{34}
     \Vec{F}_e^{(NC)}(r)=-\frac{4\pi k_BT}{\lambda_{th}^2(1+\frac{\theta^2}{\lambda_{th}^4})}  \frac{r}{1\pm(1+\frac{\theta^2}{\lambda_{th}^4})e^{\frac{2\pi r^2}{\lambda_{th}^2(1+\frac{\theta^2}{\lambda_{th}^4})}}}\hat{e}_r~.
 \end{equation}
Reassuringly the above result agrees with eq.\eqref{19} in the limit $\theta\to 0$.\\
\begin{figure}[h]
    \centering
    \includegraphics[width=0.5\textwidth]{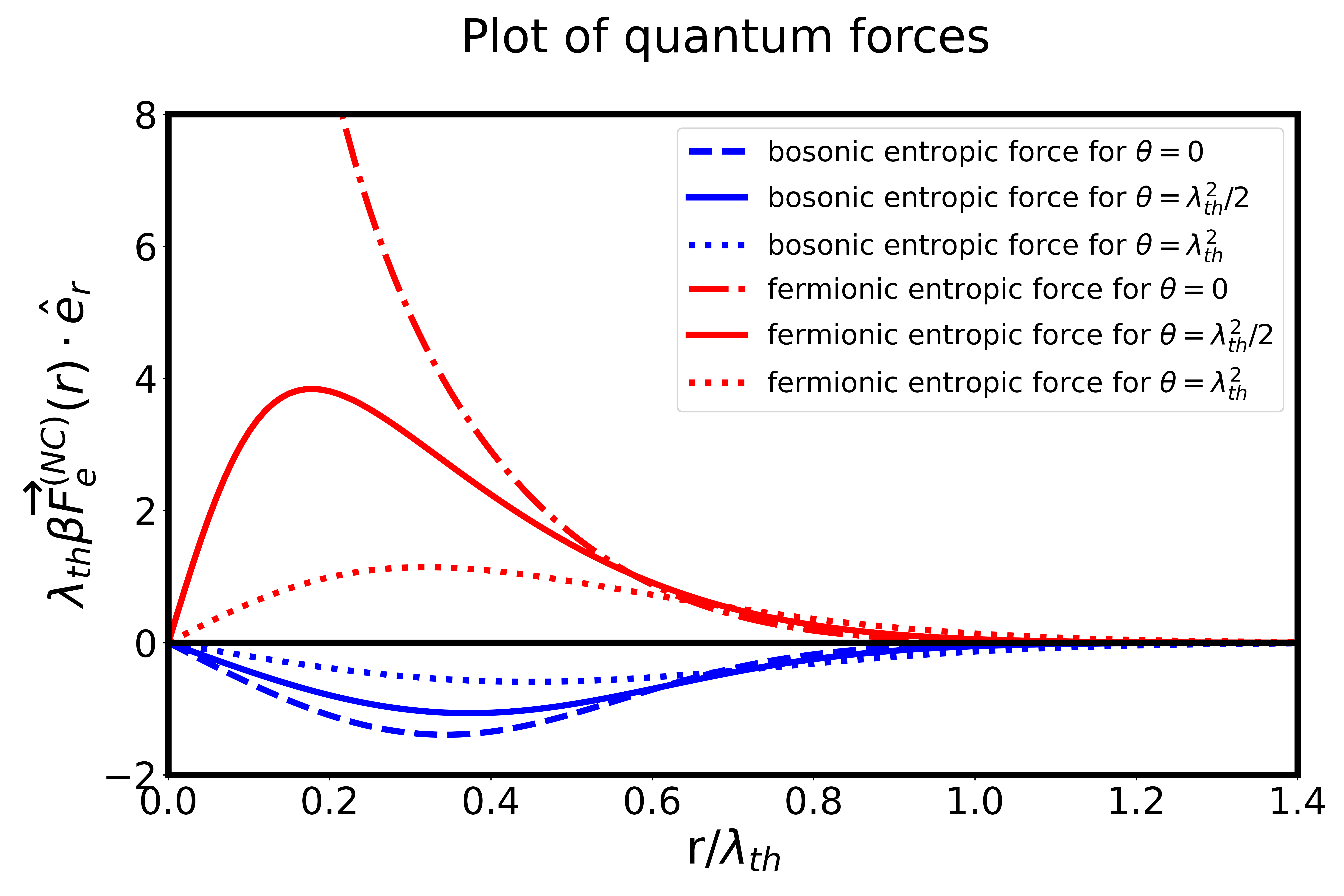}
    \caption{Interparticle entropic force in noncommutative space (as $\lambda_{th}\beta \vec{F}_e^q(r)\cdot\hat{e}_r$) is plotted against the separation (as $r/\lambda_{th}$) for bosons and fermions for different noncommutative parameters.}
    \label{fig:my_label}
\end{figure}
\noindent It is interesting to see that for fermions $\Vec{F}_e^{(NC)}(r)=0$ in the limit $r\to 0$. The vanishing of the entropic force reveals the violation of the Pauli exclusion principle in the noncommutative scenario. Similar violations were also reported in \cite{balachandran2010non},\cite{addazi2018testing}\cite{addazi2020tests}.\\
Now with the similar treatment as we did in section \ref{sub1}, we find that for bosons (in noncommutative space) at low temperature, the entropic force is given by
 \begin{equation}\label{addition3}
     \Vec{F}_e^{(NC)}(r)\approx -\frac{4\pi k_BT r}{\lambda_{th}^2(1+\frac{\theta^2}{\lambda_{th}^4})(2+\frac{\theta^2}{\lambda_{th}^4})}\hat{e}_r=-a_1\Vec{r}
 \end{equation}
 where $a_1=\frac{4\pi k_BT}{\lambda_{th}^2(1+\frac{\theta^2}{\lambda_{th}^4})(2+\frac{\theta^2}{\lambda_{th}^4})}$. This result matches with the result of eq.\eqref{20} with a noncommutative parameter modified restoration constant. For $\lambda_{th}>>\sqrt{\theta}$, that is, in the very low temperature limit there is virtually no deviation from the commutative result.\\
 For fermions, the entropic force takes the following form in the low temperature limit
\begin{equation}\label{addition3}
     \Vec{F}_e^{(NC)}(r)\approx \frac{2k_BT}{1+\frac{\theta^2}{\lambda_{th}^4}}\frac{r}{r^2+\frac{\theta^2}{2\pi\lambda_{th}^2}}\hat{e}_r~.
 \end{equation}
This result clearly shows a violation of Pauli's exclusion principle as two fermions with no separation between them feel no force. Hence, they can now sit on top of each other. However, the repulsive nature of the entropic force is still intact when the separation is non-zero.




\section{Conclusions}\label{sec4}

\noindent In this paper, we give a very general derivation of the Einstein relation of osmotic force and show its connection to the entropic force starting from the definition of Shannon entropy extended to continuous variables. We then extend our work to the quantum distribution of particles. We first consider the two particle pair correlation function which leads to a definition of the statistical potential. We show that an entropic force can be defined in this case also. We then obtain explicit expressions for this entropic force for bosons and fermions. The low temperature limit of this force for bosons gives rise to the Hooke's law. The result clearly reveals the role played by the entropic force in the formation of a Bose-Einstein condensate. The low temperature limit of the entropic force for fermions reveals Pauli's exclusion principle and also gives rise to the Neumann's radial force. We then discuss the classical limit of the entropic force between quantum particles. Our findings reveal that bosonic and fermionic particles are not independent even when there is no interaction between them. Finally, we calculate the quantum entropic force in noncommutative space. Here we find that the entropic force for fermions vanish when the separation between them is zero. This clearly reveals a violation of Pauli exclusion principle.

\bibliographystyle{ieeetr}
\bibliography{ref}

\end{document}